# Search Strategy Formulation for Systematic Reviews: issues, challenges and opportunities

Andrew MacFarlane[a], Tony Russell-Rose[b], Farhad Shokraneh[c]

[a] Centre for HCI Design, Department of Computer Science, City, University of London, Northampton Square, London EC1V 0HB, UK. Email: andym@city.ac.uk [Corresponding author].

[b] Department of Computing, Goldsmiths, University of London, New Cross, London SE14 6NW, UK. Email: T.Russell-Rose@gold.ac.uk

[c] King's Technology Evaluation Centre (KiTEC), King's College London. School of Biomedical Engineering & Imaging Sciences, 5th Floor, Becket House, 1 Lambeth Palace Road, London SE1 7EU, UK. Email: farhadshokraneh@gmail.com

Abstract
Systematic literature reviews play a vital role in identifying the best available evidence for health and social care policy. The resources required to produce systematic reviews can be significant, and a key to the success of any review is the search strategy used to identify relevant literature. However, the methods used to construct search strategies can be complex, time consuming, resource intensive and error prone. In this review, we examine the state of the art in resolving complex structured information needs, focusing primarily on the healthcare context. We analyse the literature to identify key challenges and issues and explore appropriate solutions and workarounds. From this analysis we propose a way forward to facilitate trust and transparency and to aid explainability, reproducibility and replicability through a set of key design principles for tools to support the development of search strategies in systematic literature reviews.

**Introduction**

Systematic literature reviews are used to identify and synthesize the best available evidence to support health and social care policies and decisions. The systematic review process starts with specification of the research question and development of a protocol (Hemingway and Brereton, 2009). This is followed by a search of the literature to identify relevant studies, screening of these studies, analysis of this evidence and then synthesis of finished review. Information retrieval techniques are essential to the process, particularly in the literature search and study screening phases. A key element is the literature search phase, since errors in this stage can have a significant impact on the overall success of the review and lead to an inaccurate, incomplete, and invalid outcome, e.g. by introducing bias in the synthesis (McGowan and Sampson, 2005). Rectifying a review once it has been completed can be costly; analogous to the problem in software engineering where errors overlooked early in the process (e.g. in the requirements stage) can be significantly more difficult to correct later.

In this review, we focus on literature searching, specifically the development of the search strategies used in systematic reviews. This is a complex process (Lefebvre et al, 2020; Cooper et al, 2018), in which the search methods and databases to be used are specified and peer reviewed. It is recommended that the peer review approve the protocol before searching is carried out. The search strategies are then planned and designed. Initial scoping searches may be carried out and the strategy is iteratively developed until the review team is satisfied that the strategy is fit for purpose. It is then applied to gather studies for the screening phase.

This process is time consuming and even a 'rapid review' can take between three to six months (Hemingway and Brereton, 2009), whilst a full scale review can take as much as 67 weeks on average (Borah et al, 2017). Relevant studies may be published in the meantime (Shojania et al, 2007), rendering some reviews out of date by the time they are published. Consequently, periodic updates of the search strategy may be required (Thomas et al, 2017; Shokraneh and Russell-Rose, 2020). The search strategies used should therefore be published (e.g. as an appendix to the review) and be fully replicable by third parties. They should also be transparent and explainable, so they can be reproduced as and when required. This is not only to minimise bias and errors, but also to reduce costs and hence reduce waste in research (Chalmers et al, 2014).

Automed solutions have the potential to help searchers to design, build and deploy complex search strategies, but they need to be able to trust these solutions provided. The aim of this paper is twofold: 1) to review current and best practice on search strategy formulation from both an evidence synthesis and information retrieval perspective, and to critically analyse the state of the art in terms of reproducibility, replicability, transparency, explainability and trust and 2) to propose a set of design principles to address identified issues. It should be noted that whilst we focus on healthcare, much of the analysis applies to other areas of professional search including patent search, legal research and recruitment (Russell-Rose et al, 2018).

The review is structured as follows. Key terms are then defined and the review scope outlined. Current practice in the development of search strategies in healthcare together with their deficiencies are reviewed. We then review various tools and approaches put forward to augment search strategy development methods and also assess their strengths and weaknesses. The issues identified in both current practice and supporting tools and approaches are used to build a framework of design principles, specifically to address the shortcomings in current approaches in the search strategy formulation process. We conclude with a summary, outlining steps forward.

**Definition of terms and Review Scope**

A number of key terms have been introduced above that require precise definition. This is particularly important where there is disagreement over what these terms actually mean. As we draw on literature from multiple domains (healthcare, information science, computer science) we need consistent definitions to provide a coherent narrative. These key terms are critical to understanding the role of search strategies in systematic review. First, there is a requirement for search strategies to be *reproducible* and *replicable*. As a consequence, strategies need to be also *transparent* and *explainable*, so that any searcher with the appropriate skills should be able to reuse a given search strategy appropriately. Second, since searchers desire maximal *control* over the system, there should be a high level of *trust* of the techniques and tools used. Searching for evidence to inform systematic reviews in general requires high recall, since omissions can have serious consequences for the validity of the outcome (e.g. clinical guidance may be based on incomplete or biased evidence). We start with *trust* and weave our way through the concepts such as *control*, *transparency* and *explainability* until we address *reproducibility* and *replicability*. Finally we define a number of barriers that can prevent the successful achievement of concepts identified thus far.

*Trust*

This is a user-focused term, and in our context it means that a searcher trusts the system presented and will use it to conduct a search to find information. The concept of 'Humans in the Loop' is important (Grames et al, 2019; Marshall and Wallace, 2019), as users feel that they need to be in *control* of the system, as opposed to the system controlling them. This issue of *control* reoccurs in many contexts, often associated with automated or AI solutions.

*Control*

Where the user has maximal control of the system, they are able to use it to solve their problem i.e. to formulate a search strategy that is both *transparent* and *explainable*. These two terms are related, but different. Control leads to trust in the system, lack of control leads to distrust (as will be seen below). An example of control in the context of search strategies is turning on/off auto term mapping or query expansion, and informing the user of the impact of any change. In our context this means that the user has maximal control (Paisley and Parker, 1965) over the design and development of the search strategy used in the systematic review.

*Transparency*

A system that is transparent provides the user with information on 'how' the method/algorithm works i.e. how something is done. The workings of the system should be interpretable (Doran et al, 2017). This is in contrast with an opaque system, where the details are hidden from the user and the 'how' cannot be derived from the information given (which is the default for many machine learning algorithms). In our context, it means that the user should be able to know how the search strategy worked e.g. the development process is deterministic and the outcome is predictable.

*Explainability*

A system that is explainable provides the user with information on 'why' the method/algorithm executed an action or made a decision i.e. why something is done. The system must be comprehensible e.g. to provide information relating input and output. In our context it means that a user should know why a search strategy produced a particular result set, e.g. by applying a clear and accurate mental model of its construction and function (Russell-Rose and MacFarlane, 2020). There is evidence that current systems are not explainable in that sense. Users are more likely to trust a system if the search strategies it produces are explainable, which in turn facilitates reproducibility and replicability.

*Reproducibility and Replicability*

The issue of reproducibility has become a major issue not only in healthcare sciences (Shokraneh, 2019) but also in information retrieval experimentation. This has been addressed in CLEF (Conference and Labs of the Evaluation forum) track specifically focused on reproducibility and replicability issues (Ferro et al, 2019). We use the ACM definitions (ACM, 2020), which are defined as (Plesser, 2018):

- <u>*Reproducibility:*</u> a different team uses the same test collection, and the results are directly comparable with another team's results.
- <u>*Replicability:*</u> a different team uses a different test collection (topics, documents, relevance assessment and results differ).

*Barriers*

Barriers to success in addressing concepts such as trust etc must be addressed, and throughout the review we highlight these through an analysis of the literature in the field. We identify three broad barriers that are impediments to the successful formulation of a search strategy (note that several barriers can apply any given issue identified):

- Formalism, i.e. the issue is intrinsic to the representation scheme (e.g. Boolean logic and document centric media)
- Platform, i.e. the issue is related to the tooling and technologies used (e.g. command-line query builders, proprietary databases, etc.)
- Community, i.e. the issue is related to the processes, practices and conventions adopted by the community.

We use these terms throughout the review to highlight the many problems that have been identified through the literature analysis, identifying the barriers that emerge through inspection of the literature.

**Search Strategy Development Methods: Current Practice**

Boolean logic is the de facto approach to structured searching in general and for healthcare systematic reviews in particular (Russell-Rose et al, 2018). There are variations on the process (Clarke et al, 2021), but in most cases it involves the following steps:

1. Subdividing a review topic/question into its constituent concepts. This can be facilitated by the use of a conceptual frameworks or templates such as PICO (Shokraneh, 2016);
2. Identifying which concepts should be represented as discrete search blocks or 'facets';
3. Collect terms for each facet, based either on natural language keywords or controlled vocabulary terms;
4. Combining the facets using Boolean and other operators to create an initial search strategy;
5. Testing the strategy to determine the number of results and an initial direction for refinement (with limited time and resources it may not be possible to screen every result). At this stage; it may also be appropriate to consider the use of published search templates or 'filters';
6. Sharing the results with team members for comments on addition, deletion or modification of terms;
7. Execution of the search and translation to the syntax of other databases;
8. Reporting and documenting of the search strategy.

The output of this process - namely the search strategy - can be time consuming and difficult to replicate. Moreover, a number of issues have been identified that limit both the transparency and explainability of current approaches. In the following sections, we provide an example Boolean search strategy and then review the shortcomings of this approach.

*Boolean Search strategies for Systematic Reviews*

The formulation of search strategies (steps 3-7 above) is based on facets that are developed from concepts (steps 1-2 above). Within each facet a set of synonyms or related terms is identified, and these are connected by applying the Boolean OR operator. A composite expression may then be formed by applying the Boolean AND operator across facets. The Boolean AND NOT operator may be used to remove unwanted concepts or terms. Extended Boolean operators can also be applied to narrow or broaden the search. For example, adjacency operators (e.g. NEXT, WITHIN, ADJ) can be applied to multiple terms to identify a collocation, e.g. "randomized" within 2 words of "controlled trial". Truncation or wildcard operators (*,?,&) can be applied to a single term to truncate a term or to identify spelling variations e.g. randomi?ed, to match one more characters in that string. Some databases allow terms to be restricted to particular fields, e.g. the term "diet" within the title field of an article ("diet[title]"). Restrictions on fields are also available e.g. "pain freq/3" where the search term 'pain' must appear at least 3 times for the document to be retrieved. There are many ways to construct a Boolean query with these operators including single line, block by block (Markey & Cochrane, 1981) and line by line (Bramer et al, 2018). The choice on constructing a Boolean query has a significant impact on transparency and explainability. Consider fig. 1 that shows an example of a complex and hard to interpret 'line by line' Boolean search strategy (Francis et al, 2015).

```
1. randomized controlled trial.pt.
2. controlled clinical trial.pt.
3. randomized.ab.
4. placebo.ab.
5. clinical trials as topic.sh.
6. randomly.ab.
7. trial.ti.
8. 1 or 2 or 3 or 4 or 5 or 6 or 7
9. (animals not (humans and animals)).sh.
10. 8 not 9
11. exp Child/
12. ADOLESCENT/
13. exp infant/
14. child hospitalized/
15. adolescent hospitalized/
16. (child$ or infant$ or toddler$ or adolescen$ or teenage$).tw.
17. or/11-16
18. Child Nutrition Sciences/
19. exp Dietary Proteins/
20. Dietary Supplements/
21. Dietetics/
22. or/18-21
23. exp Infant, Newborn/
24. exp Overweight/
25. exp Eating Disorders/
26. Athletes/
27. exp Sports/
28. exp Pregnancy/
29. exp Viruses/
30. (newborn$ or obes$ or "eating disorder$" or pregnan$ or childbirth or virus$ or influenza).tw.
31. or/23-30
32. 10 and 17 and 22
33. 32 not 31
```

FIG. 1. An example search strategy for Review of 'Oral protein calorie supplementation for children with chronic disease'

This example highlights two shortcomings of the Boolean approach (Russell-Rose and Shokraneh, 2020). First, it is difficult to establish the overall structure of the strategy, i.e. how facets are related and how terms are combined to form those facets. Second, such strategies do not scale well – an information need may require the use of many facets and many terms spread over a number of pages. This is a key *formalism* barrier. Although Boolean logic remains the default approach, this lack of transparency inhibits reproducibility, replicability and explainability. Additional issues regarding development methods, errors and system heterogeneity are outlined below.

*Search strategy development*

Although the basics of Boolean search can be learnt in relatively short order (MacFarlane and Russell-Rose, 2016), there are as yet no standards for designing and implementing such searches. This means that developing professional search skills entails a steep learning curve (Yoo & Mosa, 2015) and the knowledge required to successfully use Boolean strategies is typically acquired in the workplace (a *community* barrier). It takes many years to acquire the knowledge required to apply Boolean logic effectively. Designing an effective strategy can be particularly difficult when searchers do not know the best tactics to use (e.g. which combinations of keywords and operators), and as a result resort to exploratory searches (Hoang and Schneider, 2018) - a *community* barrier. This entails multiple iterations and can add further costs to the development of a systematic review (Hemmingway and Brereton, 2009; Borah et al, 2017), and it is often hard for the searcher to know when to stop (Booth, 2010) i.e. to recognize when sufficient studies been identified - both *community* and *platform* barriers. Guidelines such as PRESS 2015 can be used to audit the strategy (McGowan et al, 2016), but these contain many open questions, are subjective (Shokraneh, 2018) and it can be very expensive to revise a strategy once it has been published - a *community* barrier. Although it is considered good practice to publish the strategy along with the review, this practice is not always observed, further compromising reproducibility (Koffel and Rethlefsen, 2016; Biocic et al, 2019). Lack of documentation for the iterative process of search strategy development (Russell-Rose and Shokraneh. 2020) is both a *formalism* and *platform* barrier. This is compounded by the lack of a central repository for strategies leading to duplication of work (Koffel and Rethlefsen, 2016; Biocic et al, 2019; Shokraneh, 2018), which is both a *community* and *platform* barrier. There is also evidence that conventional professional search systems based on the Boolean model do not fully support searcher needs (Russell-Rose and Chamberlain, 2017), particularly regarding advanced functionality such as merging search queries and search results and the ability to publish search strategies to assist reproducibility and replicability - key *platform* barriers.

*Errors in search strategies*

In one study of published MEDLINE search strategies, it was shown that as many as 90% had at least one error and of these 80% were errors that limited the effectiveness of the search (Sampson and McGowan, 2006). In an investigation of Cochrane reviews, 73% were found to have a faulty search strategy design with 53% containing errors limiting the accuracy of the searches (Franco et al, 2018), with a more recent investigation report an error rate of 92.7% in search strategies for systematic reviews (Salvador-Oliván et al, 2019). Analysis of these errors shows they can be split into three types (MacFarlane and Russell-Rose, 2016):

- Strategic errors: Incorrect line number referrals (see fig. 1) can lead to invalid intermediate set merges. Using overlapping search elements leading to redundancy and increased run time for searches. Search strategy is not correctly translated to the database.
- Tactical Errors: typical are spelling errors or a missed spelling variant by incorrect use of truncation operators. Incorrect or irrelevant use of subject heading terms. Missing synonyms.
- Logical Errors: incorrect use of operator e.g. using OR when AND was required.

These errors are present in strategies developed by searchers with significant prior knowledge and experience of Boolean methods, which further highlights the lack of transparency and explainability. This is a key *formalism* barrier.

*Heterogeneity in systems*

In domains such as healthcare, it is common for users to search multiple databases. Consequently, they must translate their strategy between systems as the operators, fields and knowledge organisation schemes can differ (Bramer et al, 2018). The standard Boolean operators (AND, OR) may have the same syntax and semantics, but the NOT operator can be misinterpreted (NOT is a Boolean unary operator but is often implemented as a binary AND NOT operator in search syntax). Moreover, proximity and truncation operators can differ significantly between database search interfaces. The use of double quotes ("") is one well known method, but variations such as WITHIN, NEAR, or ADJ and NEXT are used to specify the number of words between two search terms, e.g. cancer WITHIN/2 treatments. A further complexity is that proximity operators can either be symmetric or non-symmetric e.g. cancer WITHIN/2 treatments may not be the same as treatments WITHIN/2 cancer if the database treats the order on the search terms as significant. Operators with identical syntax can be treated differently e.g. NEAR/5 in Web of Science specifies a maximum of 5 words between search terms, but the same expression in Embase specifies a maximum of 4 words. Some databases support the use of adjacency between bracketed terms and some either do not or malfunction. Furthermore, there can be a bewildering array of truncation or wildcard operators to capture a single concept, which can operate on single characters, e.g. randomi?ed or multiple characters, e.g. random*. Truncation operators can also differ in semantics e.g. '?' in Embase references a single character, whilst in Ovid this is 0 or 1 characters; confusingly '$' in Embase references 0 or 1 characters whilst in Ovid it is an unlimited number of characters. Operators can be prefixed (barely provided by existing search systems), infixed or postfixed and may apply zero or more, or one or more characters.

A further issue is that different terms can be used for the same concept in different databases, with 'AND' between fields e.g. Breast Neoplasms/ OR (Breast/ AND Neoplasms/) (Hoang and Schneider, 2018). This adds further difficulty when among vendor databases, meaning that searchers have to learn different variants of operators and knowledge organisation schemes for the same information need. This has significant implications for the reproducibility and replicability of search strategies, and requires extra effort and a complex methodology to translate between different databases (Bramer et al, 2018). This heterogeneity issue is a key *platform* barrier.

*Summary of problems with Boolean Strategies*

The complexity of Boolean search strategies leads to a lack of transparency, and it is difficult for searchers to develop effective strategies even with significant prior knowledge and experience. Search strategies can be difficult to conceptualise, time consuming to maintain and error-prone. A further complication is caused by system heterogeneity, whereby strategies need to be translated to systems with different underlying syntax, semantics and knowledge organisation schemes. Explainability is therefore limited, along with reproducibility and replicability. However, alternative methods have not managed to displace the conventional Boolean approach. Hjørland (2015) mounts a strong defence of Boolean logic and the continued role of expert searchers in complex information needs. In particular he highlights the issue of control, with exact match models such as Boolean logic providing much more control and transparency in query formulation than the best match systems such as ranking schemes. Despite their shortcomings, Boolean strategies are well known and trusted and continue to be used as the foundation for systematic reviews for healthcare and for other domains such as law, patents and recruitment. In the next section, we review approaches that have been used to augment current search strategy development to address issues raised above.

**Tools and Approaches to Augment Search Strategy Development**

A number of tools and approaches have been proposed to either augment or replace the methods outlined above for the development of search strategies. Whilst they are gaining traction they have yet to attract widespread use. Attempts to move beyond the Boolean method have been subject to similar issues regarding transparency, and unless significant benefits are demonstrated, searchers will remain reluctant to move away from conventional methods. In this section, we review a variety of 1) deployed tools and 2) general approaches including text and data extraction, automatic query expansion (AQE), Tool Automation, Hybrid schemes, Ranking and machine learning as well as potential methods in data modelling/theory and structure/representation.

*Tools 1: Text and Data Extraction*

This is an analytic approach in which data or text is extracted from documents e.g. to inform search strategies that have long been recognised as being useful tools (Thomas et al, 2011). They are used throughout the systematic review production process, but could be used in the scoping search that needs to be carried out for any review. Searchers have expressed concern about the time consuming nature of extracting information and data from documents (Hoang and Schneider, 2018), and these methods could help with some of the complexity associated with the information need and address common errors such as spelling and non-identification of synonyms. Examples include use of the PICO facet analysis scheme as a template for information extraction (Burri, 2019; Begert et al, 2020). A gold study or a set of gold studies can be analysed to extract terms, and each element of the scheme is populated using appropriate search terms connected via Boolean OR. Keyword extraction algorithms can be applied to protocols to identify useful terms (Alharbi et al, 2018). A similar approach extracting data such as the given studies data sample size could also be used to populate the search e.g. as a filter (Marshall and Wallace, 2019). Similarly keyword co-occurrence networks can be generated from analysed documents to select appropriate terms for Boolean searches (Grames et al, 2019), where some level of reproducibility is possible. Hausner et al (2012) outline a process to develop filters for queries based on terms extracted from cited documents. O'Mara-Eves et al (2015), reviewed text mining methods used to support systematic reviews and found that there is almost no attempt at replication between studies - both a *formalism* and *platform* barrier. The best approaches for the application of these techniques are therefore unknown. One attempt at reproducibility (Olorisade et al, 2017) demonstrated severe limitations from published work to match study outcomes independently. Whilst the methods can tackle some issues with the identification of search terms, they cannot address the inherent problems in the complexity of search strategies. The methods rely on Natural Language Processing technologies that may not be sufficiently transparent or explainable - a barrier to *community* acceptance. In general however, information professionals have been prepared to adopt solutions developed in collaboration with their community, such as the use of text mining techniques to identify terms to use in search strategy development, e.g. the Yale MeSH Analyzer (Yale, 2021).

*Tools 2: Automatic Query Expansion (AQE)*

Tsafnat et al (2014) claim that decision support systems (DSS) for automating search strategies can suggest tactics such as choice of keywords and operators and their combinations. Extraction methods outlined in above could also be used to populate the DSS using PICO or other frameworks. These technologies have been used in healthcare (O'Sullivan et al, 2010; 2013), but are focused on the needs of the end users (e.g. clinical staff) and are not utilized in the production of systematic reviews. DSS rely on AI technologies that limit their transparency, and are therefore not particularly explainable.

More conventional information retrieval approaches have been attempted. Scells et al (2018a; 2018b; 2019; 2020a; 2020c) have carried out work on query support for systematic reviews. This includes the transformation of a query to create a more effective one, using AQE as part of the strategy. Transformations include syntactic methods such as logical operator replacement, MeSH explosion, field restriction and adjacency replacement (Scells et al, 2018a). Scells et al (2019;2020c) built on this with semantic methods, but the underlying technology is still Boolean. Scells et al (2020a) looked at building a search strategy in an objective way by reformulating the query by looking at using gold studies to optimise the query, which can be used later in the search strategy development lifecycle. Kim et al (2011) look at the use of pseudo relevance feedback to build a decision tree to generate Boolean queries, effectively learning the best Boolean query based on traversing and testing various candidates in the tree. These schemes all rely on Boolean logic, and some of the same limitations apply together with associated barriers already identified, but with an added issue of lack of transparency.

*Tools 3: Tool Automation*

A wide variety of tools to automate the process of developing systematic reviews, including the development of search strategies, examples of which are given above (Lau, 2019). A number of key limitations to the deployment of automated tools have been identified. The systematic review production process specifies a rigorous and clearly defined framework that all searchers must adhere to as part of the development team. A key drawback is that automated tools lack compatibility with the systematic review workflow (van Altena et al, 2019; O'Connor, 2019) - a *community* and *platform* barrier. Issues to do with actually applying the tools in practice also demonstrate limitations. There is evidence of resistance to automated tools for developing search strategies when they are poorly supported (van Altena et al, 2019) - a *platform* barrier. Searchers will find it difficult to use an automated tool that has a steep learning curve associated with it (van Altena et al, 2019; Yoo & Mosa, 2015) - a *community* barrier. Usability of automated tools is questionable as they are often difficult to use, and lack the functionality to support complex information needs (van Altena et al, 2019) - a *platform* barrier. Given the above, there is clear evidence that there are considerable barriers to the adoption of automated tools. Any new system must be seamless and easier to use either to replace or augment current methods than one based on Boolean logic. Any new system must at the least be logically equivalent if not provide clear benefits over current approaches. Automated tools lack transparency and hence explainability. This leads to a lack of trust of new automated tools to support the development of search strategies (O'Connor et al, 2019). The evidence is that these tools have all of the drawbacks but none of the benefits of conventional Boolean search strategy development, which in part explains their limited adoption.

*Tools 4: Ranking algorithms, Machine Learning and Learning to Rank*

Ranking algorithms have been around for many years and are routinely used in many contexts such as web search to present relevant results to users. These are less used in professional search due in part to a perception of reduced transparency in the algorithms (Russell-Rose et al, 2018). These ranking algorithms have been extended to classification and learning to rank schemes. Standard ranking models such as BM25 have been augmented with learning to rank schemes (Li et al, 2018). Machine learning has been deployed to automatically filter articles for a given review, using NLP methods to identify terms (Burri, 2019) and others have applied learning to rank methods on those terms extracted from a protocol (Minas et al, 2018). Neural Networks have also been used to find articles relevant to a systematic review topic (Marshall and Wallace, 2019). Scells et al (2020b) suggest methods for using Learning to Rank models to rank Boolean queries in much the same vein as the AQE methods reviewed above. Dynamic machine learning approaches have also been proposed (Cormack and Grossman, 2018). Thomas et al (2017) review methods for living systematic reviews, using information filtering methods that include machine learning for classification and information extraction that require periodic updates. These methods can and do show utility, but suffer from a major problem. As with many AI technologies their explainability is limited, and this has undoubtedly inhibited their adoption, and all the barriers identified in the previous sections apply here. Further issues with such technologies include uses on very specific problems e.g. one type of study design or studies within one key topic such as randomised controlled trials. They are often not usable for multi-faceted searches of the type addressed in this review, but can be used as a useful tool for either for query filters ("Hedges") or query reformulation once the query has been developed (Del Fiol et al, 2018; Russell-Rose et al, 2021) and some models such as BioBERT are beginning to have a significant impact in the field (Lee et al, 2020).

*Tools 5: Hybrid Schemes*

Karimi et al (2010) undertook a number of different experiments, in particular a hybrid scheme where Boolean and rankings schemes are used in conjunction. They focus very much on the reproducibility issue and the problems formulating search strategies using Boolean logic, but are realistic as to the downsides of ranking. There is evidence that shows that the hybrid method does push more relevant documents higher up the ranked list, but this still does not address the conceptual problem of building the search strategy in the first place.

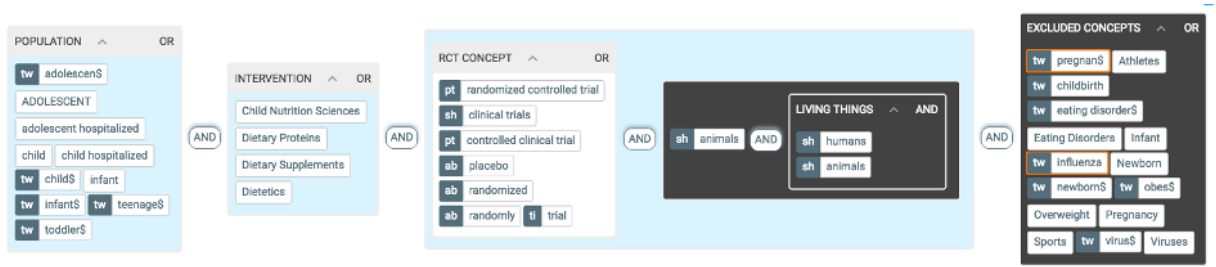

FIG. 2. 2Dsearch visual block representation of line by line strategy from FIG. 1.

*Tools 6: Structures/Representation*

It is possible to move away from the imperative methods used in Boolean search strategies to a more declarative method where relevant documents are specified by what is wanted, rather than how to obtain it. Verberne (2018) argues for transparency in personalised professional search using knowledge graphs to create a representation of the concepts the searcher is looking for. Russell-Rose et al (2019) and Russell-Rose (2019) outline a visual approach where a 2D sketchpad creates a canvas where blocks (concepts) can be created and relationships established which mirrors the Boolean constructs, but with a more direct mapping between the visual representation and the intended semantics (see fig. 2, a visual representation of the search strategy presented in fig. 1). Other visual methods have been tried before, which are reviewed by Russell-Rose et al (2019) e.g. Venn diagrams and 'dust and magnet' representations. These methods still rely on Boolean logic, but provide some level of transparency and better support for concept representation and provide a partial solution to the problem, addressing in part the *formalism* barrier e.g. by supporting the separation of search strategies from the platform when recording and sharing it.

*Approaches 1: Data Modelling/Theory*

The underlying theory used in systematic reviews is Boolean logic, with extensions to support proximity (a special case of AND), truncation/wildcards (a special case of OR) and search fields (based on metadata). There are many potential ways to deal with this, either replacing the logic with something new, or abstracting the details of the logic away and representing it visually (see above). Ranking as an alternative logic could be used, but is not accepted by the community (see above). There are logics such as 'fuzzy' logic, but these have not gained much headway in the IR field generally (Robertson, 1978), and are likely to face the same resistance to implementation as ranking schemes for many of the same reasons. No serious attempt has been made to use different models and theories apart from those reviewed above to the best of our knowledge.

*Approaches 2: Explainability*

There is little work on explanations for search strategies used in systematic reviews. Much work concentrates on the screening process e.g. Mi & Jiang (2019) and Thomas et al (2019), in particular the interpretability of those results. This does not really deal with the essential problem in professional search where the issue is earlier in the process. Thomas et al (2019) focuses on web searches, not professional ones. Apart from this, there does not appear to be much in the way of primary research being done specifically on professional search, apart from Russell-Rose and MacFarlane (2020) and Verberne (2018), that both identify gaps and issues to be addressed.

*Summary of Tools and Approaches for Search Strategy Development*

In summary, there have been attempts to address the shortcomings of conventional approaches or augment them. Such attempts rely either on Boolean approaches and share many of their shortcomings, or on machine learning and/or information extraction on top of standard information retrieval models (e.g. BM25). Some of the methods can be used to supplement Boolean approaches and improve the search strategy development, but do not address the problem of developing the initial query as they only become useful in a second stage for query reformulation from identified relevant or 'gold standard' studies. Some tools can be used to develop an initial query for non-complex topics i.e. Unsilo (2021) and ResearchRabbit (2021), but deploying the tools to address complex topics given the inherent tact knowledge required in the same way is a long way off, if possible at all.

There is little work in actually understanding the underlying problem in terms of frameworks which are explainable/transparent or conceptual ideas which would allow the searcher to better understand their information needs. In some cases, structure (such as the use of PICO, SPIDER, or SPICE) is used. There is resistance to the introduction of automated tools that might assist the searchers (van Altena et al, 2019; O'Connor, 2019; Yoo & Mosa, 2015), due to lack of compatibility with workflows and steep learning curves. Approaches examined in this section therefore retain many of the same limitations, and many lack transparency and explainability leading to an erosion of trust. Efforts to address the limitations of Boolean methods are laudable, but we need new solutions to ensure that search strategies are transparent and trustworthy, and facilitate explainability, reproducibility and replicability.

**Discussion: A Framework for Search Strategy Formulation**

Boolean search strategies have been used successfully for many years to resolve complex information needs in healthcare and other domains such as legal research, patent search and recruitment. Professional searchers have built up considerable tacit knowledge to ensure that relevant studies are identified for evidence synthesis, but there are still significant barriers regarding reproducibility and replicability. There is some level of transparency and explainability which whilst limited does facilitate system control and hence trust. However, current methods can be hard to learn, difficult to conceptualise, time consuming to maintain and error-prone, with the heterogeneity of systems adding further complexity. Laudable attempts to address this problem have been made, but many rely on AI and machine learning technologies that also lack transparency and explainability. In addition, many alternative toolsets and approaches are difficult to use, have a steep learning curve and are not easily integrated into the systematic review production workflow. They therefore demonstrate the same drawbacks as conventional methods, but few of the advantages particularly in terms of trust. The community needs to develop new conceptual frameworks that offer a better way to resolve complex structured information needs in the evidence synthesis process. These frameworks need to help the user better articulate the structure of their information needs, to aid explainability and transparency, and to be more scalable to aid reproducibility and replicability. In this paper we have identified many problems with current methods, and propose a framework that sets down key principles to offer solutions to these problems.

This framework is outlined in the following set of design principles and a mapping from the barriers identified above to those principles. Our aspiration is that this framework will inform the development of a new conceptual framework for search strategy formulation (Russell-Rose and MacFarlane, 2020). The five design principles are as follows:

1. Provide support for transparency in the mapping between logical structure and physical structure, using visual representations to communicate conceptual structure and relationships and encourage exploration.
2. Adopt scalable, declarative formalisms that accommodate complexity and support abstraction and encapsulation, e.g. allowing users to switch between overview and detail views and independently manipulate individual query elements
3. Delegate lower-level syntactic operators to system functions, e.g. replacing error-prone string manipulation with automated translation to different query syntaxes.
4. Provide real time feedback on query effectiveness, allowing users to perceive the contribution of individual query elements and understand how to make queries more effective.
5. Provide support for collaboration and team working, e.g. through repositories of best practice examples that facilitate versioning, sharing and peer review.

Table 1 provides a mapping between the barriers identified in this review (column 2), their source (column 3), potential solutions in the form of the relevant design principles (column 4) and the barrier category (column 5). In the last column we categorise issues according to the stage in the search strategy formulation process to which they apply (as defined above).

Table 1 - Summary of problems identified and potential solutions, issue type and process stage.

| # | Issue | Citation | Solution (Design principle) | Barrier type | Process stages (section 4) |
|---|---|---|---|---|---|
| 1 | Difficulty in identifying and/or articulating strategy structure. | Russell-Rose and Shokraneh (2020). | Support transparency in mapping (P1). | Formalism | Planning/designing: 1,2,3. |
| 2 | Lack of scalability of Boolean strings. | Russell-Rose and Shokraneh (2020). | Adopt Scalable mechanisms (P2). | Formalism | Testing/Running: 4,5,7. |
| 3 | Steep learning curve. | Yoo & Mosa (2015), van Altena et al (2019). | Increase transparency, reduce complexity (P1-4). | Platform/ community | All: 1-8. |
| 4 | Lack of design standards for search strategy development. | Hoang and Schneider (2018). | Help community develop design standards (P1-5). | Community | Planning/designing: 1,2,3. |
| 5 | Costs of development (time, resources, budgets). | Hemmingway and Brereton, (2009), Borah et al, (2017). | Reduce costs by applying all EARS principles, making the process faster (P1-5). | Platform/ community | All: 1-8. |
| 6 | Lack of comprehensive development guidelines. | Shokraneh (2018). | Revise in line with design standards (P1-5). | Community | Planning/designing: 1,2,3. |
| 7 | Lack of a central repository for sharing strategies and/or results (standard templates & format). | Koffel and Rethlefsen (2016); Biocic et al. (2019) Shokraneh (2018). | Provide support for collaboration/team working (P5). | Platform/ community | Reporting/sharing: 6,8. |
| 8 | Limited functionality for search management. | Russell-Rose & Chamberlain (2017), van Altena et al (2019). | Abstractions are required to provide a variety of functions (P2-4). | Platform | All: 1-8. |
| 9 | Prevalence of errors in published strategies. | Sampson and McGowan (2006); Franco et al. (2018). | Eliminate errors via abstractions, provide feedback on user queries (P1-4). | Formalism | Testing/Running: 4,5,7. |
| 10 | Inefficiencies due to system heterogeneity & need for translation. | Bramer et al. (2018). | Standard search strategy model required (P1-4). | Platform | Testing/Running: 4,5,7. |
| 11 | Lack of replication in automated systems. | O'Mara-Eves et al. (2015) | Increase transparency, abstract complexity out (P1-4). | Formalism/ Platform | Testing/Running: 4,5,7. |
| 12 | Limited compatibility with existing workflows. | van Altena et al (2019), O'Connor (2019). | User acceptance of Design principles (P1-5). | Platform/ community | All. |
| 13 | Poor support for query management. | van Altena et al (2019). | In built support for users e.g. real time feedback on query effectiveness (P4). | Platform | Testing/Running: 4,5,7. |
| 14 | Lack of usability & accessibility of existing platforms. | van Altena et al (2019). | Increase transparency, abstract away complexity (P1-4). | Platform | Testing/Running: 4,5,7. |
| 15 | Lack of documentation for iterative process of designing the search. | Russell-Rose and Shokraneh (2020). | Provide support for collaboration/team working (P5). | Formalism/ Platform | Planning/designing: 1,2,3. |
| 16 | Separation of search strategies from platform during recording / reporting/sharing. | Russell-Rose Chamberlain and Shokraneh (2019). | Provide support for collaboration/team working (P5). | Formalism | Reporting/sharing: 6,8. |

| 17 | Out of date and unreliable search strategies (need for living search strategies). | Thomas at al. (2017), Shojania et al, (2007) | Provide support for collaboration/team working (P5). | Formalism/Platform | Reporting/sharing: 6,8. |

In applying these principles it should be noted that searchers' needs are a continuum from those that are relatively simple and machine applicable (can be automated) such as a scoping review, to those that are relatively complex and require significant human input, such as a full systematic review. Finding the right way to apply the principles to these different types of needs is important to engender trust in any implemented system that embodies such principles. The key to gaining trust is to enhance explainability, by providing more transparency for search strategy development, which in turn will enhance replicability and reproducibility.

**Summary**

In this paper we have used the key concepts of reproducibility, replicability, transparency, explainability and trust to examine the problems professional searchers have in developing search strategies that are effective in identifying relevant literature for evidence synthesis. We have focused on healthcare, but other domains such as patents, law and recruitment stand to gain from any improvement that can be derived by addressing these key concepts. The contribution of this paper is a framework with which the IR and systematic review community can use to address the many problems that have been identified. It is our ambition to bring the communities together and promote collaboration between them to address the problems we identify and to apply the framework to generate solutions to those problems.